\documentclass[conference]{IEEEtran}
\usepackage{acronym}
\usepackage{amsmath,amssymb,amsfonts}
\usepackage{algorithmic}
\usepackage{graphicx}
\usepackage{textcomp}
\usepackage{xcolor}
\usepackage{etoolbox}
\usepackage{slantsc}
\usepackage{xspace}
\usepackage[abbreviations]{foreign}
\usepackage{tikz}
\usepackage{pgfplots}
\usepackage[binary-units]{siunitx}
\usepackage[eulergreek]{sansmath}
\usepackage[ruled,vlined,linesnumbered]{algorithm2e}
\usepackage[most]{tcolorbox}
\usepackage{booktabs}
\usepackage[inline]{enumitem}
\usepackage[backend=bibtex,sorting=none,style=ieee,isbn=false]{biblatex}
\usepackage{multicol}
\usepackage[hidelinks]{hyperref}
\usepackage{cleveref}
\usepackage{color, colortbl}
\usepackage{flushend}
\definecolor{Gray}{gray}{0.9}
\def\BibTeX{{\rm B\kern-.05em{\sc i\kern-.025em b}\kern-.08em
T\kern-.1667em\lower.7ex\hbox{E}\kern-.125emX}}

\makeatletter
\newif\ifAC@uppercase@first%
\def\Aclp#1{\AC@uppercase@firsttrue\aclp{#1}\AC@uppercase@firstfalse}%
\def\AC@aclp#1{%
  \ifcsname fn@#1@PL\endcsname%
    \ifAC@uppercase@first%
      \expandafter\expandafter\expandafter\MakeUppercase\csname fn@#1@PL\endcsname%
    \else%
      \csname fn@#1@PL\endcsname%
    \fi%
  \else%
    \AC@acl{#1}s%
  \fi%
}%
\def\Acp#1{\AC@uppercase@firsttrue\acp{#1}\AC@uppercase@firstfalse}%
\def\AC@acp#1{%
  \ifcsname fn@#1@PL\endcsname%
    \ifAC@uppercase@first%
      \expandafter\expandafter\expandafter\MakeUppercase\csname fn@#1@PL\endcsname%
    \else%
      \csname fn@#1@PL\endcsname%
    \fi%
  \else%
    \AC@ac{#1}s%
  \fi%
}%
\def\Acfp#1{\AC@uppercase@firsttrue\acfp{#1}\AC@uppercase@firstfalse}%
\def\AC@acfp#1{%
  \ifcsname fn@#1@PL\endcsname%
    \ifAC@uppercase@first%
      \expandafter\expandafter\expandafter\MakeUppercase\csname fn@#1@PL\endcsname%
    \else%
      \csname fn@#1@PL\endcsname%
    \fi%
  \else%
    \AC@acf{#1}s%
  \fi%
}%
\def\Acsp#1{\AC@uppercase@firsttrue\acsp{#1}\AC@uppercase@firstfalse}%
\def\AC@acsp#1{%
  \ifcsname fn@#1@PL\endcsname%
    \ifAC@uppercase@first%
      \expandafter\expandafter\expandafter\MakeUppercase\csname fn@#1@PL\endcsname%
    \else%
      \csname fn@#1@PL\endcsname%
    \fi%
  \else%
    \AC@acs{#1}s%
  \fi%
}%
\edef\AC@uppercase@write{\string\ifAC@uppercase@first\string\expandafter\string\MakeUppercase\string\fi\space}%
\def\AC@acrodef#1[#2]#3{%
  \@bsphack%
  \protected@write\@auxout{}{%
    \string\newacro{#1}[#2]{\AC@uppercase@write #3}%
  }\@esphack%
}%
\def\Acl#1{\AC@uppercase@firsttrue\acl{#1}\AC@uppercase@firstfalse}
\def\Acf#1{\AC@uppercase@firsttrue\acf{#1}\AC@uppercase@firstfalse}
\def\Ac#1{\AC@uppercase@firsttrue\ac{#1}\AC@uppercase@firstfalse}
\def\Acs#1{\AC@uppercase@firsttrue\acs{#1}\AC@uppercase@firstfalse}
\robustify\Aclp
\robustify\Acfp
\robustify\Acp
\robustify\Acsp
\robustify\Acl
\robustify\Acf
\robustify\Ac
\robustify\Acs
\def\AC@@acro#1[#2]#3{%
  \ifAC@nolist%
  \else%
  \ifAC@printonlyused%
    \expandafter\ifx\csname acused@#1\endcsname\AC@used%
       \item[\protect\AC@hypertarget{#1}{\acsfont{#2}}] #3%
          \ifAC@withpage%
            \expandafter\ifx\csname r@acro:#1\endcsname\relax%
               \PackageInfo{acronym}{%
                 Acronym #1 used in text but not spelled out in
                 full in text}%
            \else%
               \dotfill\pageref{acro:#1}%
            \fi\\%
          \fi%
    \fi%
 \else%
    \item[\protect\AC@hypertarget{#1}{\acsfont{#2}}] #3%
 \fi%
 \fi%
 \begingroup
    \def\acroextra##1{}%
    \@bsphack
    \protected@write\@auxout{}%
       {\string\newacro{#1}[\string\AC@hyperlink{#1}{#2}]{\AC@uppercase@write #3}}%
    \@esphack
  \endgroup}
\makeatother
\acrodef{DCAP}{data  center  attestation  primitives}
\acrodef{DNN}{deep neural network}
\acrodef{DP}{differential privacy}
\acrodef{CF}{collaborative filtering}
\acrodef{EPC}{enclave page cache}
\acrodef{ECDH}{elliptic-curve Diffie–Hellman}
\acrodef{ECDSA}{elliptic curve digital signature algorithm}
\acrodef{EPID}{enhanced  privacy identifier}
\acrodef{HE}{homomorphic encryption}
\acrodef{IAS}{Intel attestation service}
\acrodef{ML}{machine learning}
\acrodef{MPC}{secure multi-party computation}
\acrodef{QE}{quoting enclave}
\acrodef{SGD}{stochastic gradient descent}
\acrodef{SGX}{software guard extensions}
\acrodef{TEE}{trusted execution environments}
\acrodef{RMSE}{root mean square error}
\acrodef{MSE}{mean square error}
\acrodef{RMW}{random model walk}
\acrodef{MF}{matrix factorization}
\acrodef{FL}{federated learning}
\acrodef{DL}[DLS]{decentralized learning systems}
\acrodef{STL}{standard template library}
\acrodef{SDK}{software development kit}
\acrodef{OS}{operating system}
\acrodef{IO}[I/O]{input and output}
\acrodef{TLB}{translation lookaside buffer}
\acrodef{TCB}{trusted computing base}
\acrodef{ocall}{outside call}
\acrodef{ecall}{enclave call}
\acrodef{RPC}{remote procedure call}
\acrodef{KNN}{K nearest neighbours}
\acrodef{TCG}{trusted computing group} 
\graphicspath{ {figures/} }

\newcommand{\rex}{\textsc{Rex}\xspace}
\newcommand{\nextnr}{\stepcounter{AlgoLine}\ShowLn}

\newboolean{showcomments}
\setboolean{showcomments}{true}
\ifthenelse{\boolean{showcomments}}
{ \newcommand{\mynote}[3]{
		\fbox{\bfseries\sffamily\scriptsize#1}
		{\small$\blacktriangleright$\textsf{\emph{\color{#3}{#2}}}$\blacktriangleleft$}}
	\newcommand{\zzz}[1]{{\setlength{\fboxsep}{2pt}\fcolorbox{black}{yellow}{\textsf{\emph{#1}}}}\xspace}}
{ \newcommand{\mynote}[3]{}
	\newcommand{\zzz}[1]{}}

\usepgfplotslibrary{units,colorbrewer,groupplots,fillbetween}
\usetikzlibrary{patterns}

\makeatletter
\newcommand{\removelatexerror}{\let\@latex@error\@gobble}
\makeatother

\let\oldnl\nl%
\newcommand{\nonl}{\renewcommand{\nl}{\let\nl\oldnl}}
\newcommand{\func}[1]{\mathtt{#1}}

\Crefname{algocf}{Algorithm}{Algorithms}                                        
\AtBeginEnvironment{algorithm}{%
	\DontPrintSemicolon                                                         
	\SetKwProg{Procedure}{Procedure}{:}{end}                                    
	\SetKwInOut{Input}{input}                                                   
	\SetKwInOut{Output}{output}                                                 
	\fontsize{8}{9.6}                                                           
} 

\makeatletter                                                                   
\renewcommand{\algocf@makecaption@ruled}[2]{%
	\global\sbox\algocf@capbox{\hskip\AlCapHSkip%
		\setlength{\hsize}{\columnwidth}%
		\addtolength{\hsize}{-2\AlCapHSkip}%
		\parbox[t]{\hsize}{\algocf@captiontext{#1}{#2}}}%
}%
\makeatother

\makeatletter
\begingroup\endlinechar=-1\relax
\everyeof{\noexpand}%
\edef\x{\endgroup\def\noexpand\homepath{%
		\@@input|"kpsewhich --var-value=HOME" }}\x
\makeatother

\def\overleafhome{/tmp}
\newcommand{\inputplot}[2]{%
	\ifx\homepath\overleafhome%
	\IfBeginWith{#1}{plots}{\includegraphics{main-figure#2.pdf}}{#1}%
	\else%
	{\sffamily\scriptsize\input{#1}}
	\fi}

\bibliography{references,refsAMK}

\makeatletter
\newcommand\notsotiny{\@setfontsize\notsotiny\@vipt\@viipt}
\makeatother

\newcommand\copyrighttext{%
    \footnotesize \textcopyright 2022 IEEE.
    Personal use of this material is permitted.
    Permission from IEEE must be obtained for all other uses,
    in any current or future media, including reprinting/republishing this
    material for advertising or promotional purposes, creating new collective
    works, for resale or redistribution to servers or
    lists, or reuse of any copyrighted component of this work in other works.
    Pre-print version. Presented on May 31st 2022 in the {36th IEEE International Parallel and Distributed Processing Symposium (IPDPS '22)}. For the final published version, please refer to DOI \href{https://doi.org/10.1109/IPDPS53621.2022.00050}{10.1109/IPDPS53621.2022.00050}.}
\newcommand\copyrightnotice{%
\begin{tikzpicture}[remember picture,overlay]
    \node[anchor=south,yshift=10pt,fill=yellow!20] at (current page.south) {\fbox{\parbox{\dimexpr\textwidth-\fboxsep-\fboxrule\relax}{\copyrighttext}}};
\end{tikzpicture}%
}

\begin{document}

\title{TEE-based decentralized recommender systems: \\The raw data sharing redemption}

\author{\IEEEauthorblockN{Akash Dhasade, Nevena Dresevic, Anne-Marie Kermarrec, Rafael Pires}
\textit{EPFL - Swiss Federal Institute of Technology}\\
Lausanne, Switzerland\\
\texttt{first.last}@epfl.ch
}

\maketitle
\copyrightnotice

\pagestyle{plain}%

\begin{abstract}
Recommenders are central in many applications today. The most effective recommendation schemes, such as those based on \ac{CF}, exploit similarities between user profiles to make recommendations, but potentially expose private data.  
Federated learning and decentralized learning systems address this by letting the data stay on user's machines to preserve privacy: each user performs the training on local data and only the model parameters are shared. However, sharing the model parameters across the network may still yield privacy breaches. 
In this paper, we present \rex{}, the first enclave-based decentralized CF recommender.
\rex{} exploits \Ac{TEE}, such as Intel \ac{SGX}, that provide shielded environments within the processor to improve convergence while preserving privacy.
Firstly, \rex{} enables raw data sharing, which ultimately speeds up convergence and reduces the network load. 
Secondly, \rex{} fully preserves privacy. We analyze the impact of raw data sharing in both \ac{DNN}  and \ac{MF} recommenders and showcase the benefits of trusted environments %
in a full-fledged implementation of \rex{}.
Our experimental results demonstrate that through raw data sharing, \rex{} significantly decreases the training time by $18.3 \times$ and the network load by 2 orders of magnitude over standard decentralized approaches that share only parameters, while fully protecting privacy by leveraging trustworthy hardware enclaves with very little overhead.
\end{abstract}

\begin{IEEEkeywords}
privacy, security, recommender systems, SGX
\end{IEEEkeywords}

\section{Introduction}
\acresetall
Recommendation systems are now central in a wide variety of web applications to help users navigate through the exponentially growing volume of data. 
They help users to pick the items they are likely to buy on online stores, 
predict which movies they are willing to watch on streaming platforms~\cite{10.1145/2843948} and
decide which information to display on social media~\cite{DBLP:conf/hpca/GuptaWWNR0CHHJL20},
 to cite a few.
While many approaches exist, \ac{CF}~\cite{su:2009:collaborative,DBLP:conf/www/DasDGR07} is arguably the most successful approach and has been  widely adopted in industry.
\ac{CF} exploits the similarities between users to learn their preferences and accurately compute recommendations or predictions. 

Precisely because recommenders learn  users' preferences, they represent a serious privacy threat. User profiles are stored on service providers that may involuntarily leak them through data breaches or voluntarily release their databases %
for commercial purposes. %
CF has to face a dilemma, typically sacrificing accuracy or efficiency to guarantee privacy.  For instance, relying on homomorphic encryption to encrypt user data provides a high level of privacy but is known to be notoriously impractical~\cite{lauter:2011:homomorphic:practical}.  Differential privacy has been applied to recommendation systems at the price of significantly hampered accuracy~\cite{zhu:2016:dp:recommender}.

More recently, \ac{FL}~\cite{mcmahan:2017:fl} and \ac{DL}~\cite{ormandi:2013:gossiplearning} 
took an orthogonal strategy: they have been introduced as an attractive alternative to address both scalability and privacy of \ac{ML} systems. In a nutshell, \ac{FL} and \ac{DL} consider a scenario  where the data is fully distributed, \ie, raw data is produced by users, who hold a single personal profile record and the model is trained locally.
Such approaches require that users' data stay where it is  produced, thus limiting their exposure.
Yet, models need to be aggregated in order to provide relevant recommendations for unseen items.
Learning tasks performed on user devices are merged on a central server in \ac{FL} or through a gossip-based protocol in \ac{DL}.  
Instead of moving raw data, these approaches only allow nodes to share processed data (\eg, weights and gradients), which unfortunately does not fully protect individuals' privacy.
It has been shown that sharing model parameters across the network reveals some information about user profiles and yields privacy breaches~\cite{geiping:2020:invertgradients,sablayrolles2019white,ganju:2018:FCNNinference}. %
Finally, sharing model parameters might induce considerable network traffic due to the large size of models.
For instance, the \ac{DNN} model that we use in our experiments has more than \num{200000} parameters that would be exchanged at each iteration of a decentralized learning training task, whereas by exchanging only a few data items per epoch we are able to achieve an equivalent test error target, as we will show later.

\textit{If only data could be shared safely in a network of machines, recommenders could achieve at once privacy, accuracy and scalability}.
In this paper, we propose \rex{}, the first %
enclave-based decentralized recommender that achieves this three-dimensional goal.
\rex{} exploits \ac{TEE}, such as Intel \ac{SGX}, that provide shielded environments within the processor.
Since users cannot inspect what is being processed inside the enclaves on their own machines, \rex{} enables raw data sharing among nodes.
Given \ac{SGX} assurances, \rex{} enables to conceal sensitive data from adversaries both during the decentralized training and the communication phases.
The benefits of enabling raw data sharing in decentralized systems with \rex{} is threefold:
\begin{enumerate*}[label={(\roman*)}]
	\item disseminating raw data in the system speeds up the training time,
	\item sharing data instead of model parameters in recommender systems, where datasets are sparse, significantly reduces network traffic in the system, and
	\item the use of \acp{TEE} fully preserves users' privacy.
\end{enumerate*}

In this paper, we make the following contributions:

\begin{itemize}
	\item We propose the design and evaluation of \rex{}~\footnote{Source code is available at \url{https://github.com/rafaelppires/rex}.}, a novel  decentralized recommender that avoids trading-off accuracy or efficiency for privacy. \rex{} relies on \acs{SGX} to quickly compute a recommendation model while limiting network bandwidth and  without sacrificing users' privacy. 
	\item We demonstrate the benefits of raw data-sharing over model sharing in recommendation systems through an extensive experimental study. We compared those approaches along both model quality and system metrics. We considered two publicly available datasets (MovieLens latest and 25M~\cite{grouplens:2021:movielens}), two different models (matrix factorization~\cite{koren2009matrix} and DNN~\cite{goodfellow:2016:deep}) as well as two network topologies (small-world~\cite{SW} and random~\cite{erdos:1959:randomgraphs}), across two decentralized learning algorithms (RMW~\cite{ormandi:2013:gossiplearning} and D-PSGD~\cite{lian:2017:dpsgd}), thus demonstrating the generality of our approach.
	\item We ran \rex{} on Intel Xeon E-2288G CPUs and demonstrate its viability on a real system. More specifically, we show that the overhead of SGX remains low.
\end{itemize}

The rest of the paper is organized as follows: We present some background on recommenders, ML models and SGX technology in Section \ref{section:background}. The design of \rex{} is described in Section \ref{section:rex}. We present an extensive experimental evaluation demonstrating the benefits of \rex{} over model sharing as well as the overhead of using SGX on a real implementation in Section \ref{section:eval}. Related work is surveyed in Section \ref{section:related} before concluding in Section \ref{sec:conclusion}. 
\section{Background}
\label{section:background}

\subsection{Personalized recommendation}

\paragraph{Collaborative filtering} %
While various approaches exist to achieve recommendation, in this paper we focus on the most popular one, namely collaborative filtering (CF) ~\cite{DBLP:journals/fthci/EkstrandRK11}.
\ac{CF}  predicts the items a user will be interested in not only from the user's own past activities but those of every other user.
We consider a set of $n$ users $U= \{u_1, u_2,\ldots,u_n\}$ and a set of $m$ items $I= \{i_1, i_2,\ldots,i_m\}$.
To each user $u \in U$ is associated a profile $P_u$, which contains the user's opinions on the items she has seen/liked/clicked in the past.
The profile $P_u$ is a collection of tuples $<i,v>$ representing the rating $v$ on item $i$ by user $u$.
Ratings may be binary or convey a particular value.
The user-item interactions can then be represented as a matrix $A \in \mathbb{R}^{n \times m}$ composed of all $n$ users and $m$ items.
Given that users only interact with a few items, the goal of \ac{CF} is to fill up the matrix with predictions for the missing values.

\paragraph{\Acf{MF}} \acs{MF}~\cite{koren2009matrix} is among the most popular approaches that decompose the user-item interaction matrix $A \in \mathbb{R}^{n \times m}$ into a product of two  matrices of lower dimension $X \in \mathbb{R}^{n \times k}$ and $Y \in \mathbb{R}^{m \times k}$ representing embeddings that summarise  user tastes and item profiles, respectively. The matrices $X$ and $Y$ can then be used to directly infer a score. Formally, a \ac{MF} objective function can be defined as 
\begin{equation*}
	J(X, Y) = \frac{1}{2}||A - XY^T||^2  = \frac{1}{2} \sum_{i = 1}^{n} \sum_{j = 1}^{m} (a_{ij} - \sum_{l = 1}^{k} x_{il}y_{jl})^2
\end{equation*}
where the matrix $XY^T$ is the rank-$k$ approximation of $A$ and the goal is to find $X$ and $Y$ that minimize the error function $J$. 
This is achieved by using some method for optimizing objective functions, like \ac{SGD}.

Often in practice, matrix $A$ is sparse where only some of the user-item interactions are known. In this case, the problem is modified to find an optimal rank-$k$ approximation for only known values of $A$. Additional regularisation terms and regularisation parameter $\lambda$ are added to stabilise the optimisation process. Also, bias vectors $b \in \mathbb{R}^{n \times 1}$ and $c \in \mathbb{R}^{m \times 1}$ are included to account for the fact that some users tend to give higher or lower ratings than others while particular items may receive higher or lower ratings. Including all terms, the loss function $J(X, Y, b, c)$ can be defined as
\begin{equation*}
	\frac{1}{2} \sum_{(i, j) \in I} (a_{ij} - b_i - c_j - \sum_{l = 1}^{k} x_{il}y_{jl})^2 + \frac{\lambda}{2} ||X||^2 + \frac{\lambda}{2} ||Y||^2
\end{equation*}
where $I$ represents the set of indices for known values in $A$. 
Upon learning matrix $X$ and matrix $Y$, the predictions $p_{ij}$ for user $i$ and item $j$ are obtained as
$p_{ij} = X_i \cdot Y_j + b_i + c_j$.

\paragraph{Deep Neural Networks} Personalized recommendation can also be accomplished with deep learning approaches (DNN)~\cite{DBLP:conf/hpca/GuptaWWNR0CHHJL20}. 
A \ac{DNN} is an artificial neural network that uses several layers of nodes with non-linear activation functions to learn complex functions which capture patterns in the input data to achieve a desired prediction. 
For the problem of predicting ratings, the data is represented as triplets of the form $<\mathrm{user}_i,\mathrm{item}_j, \mathrm{rating}>$. %
We add an intermediate embedding layer which can be considered equivalent to the lower-rank matrices %
described in the MF section above.
Each pair $<\mathrm{user}_i, \mathrm{item}_j>$ indexes corresponding embeddings %
in matrices $X$ and $Y$, which are concatenated and fed as input to the \ac{DNN}.
Its output, in turn, is unidimensional, representing the predicted rating for the given combination.
The learning process occurs on both the weights of the neural network and the embedding matrices.
Finally, learnt embeddings are used to predict ratings for unseen user-item pairs.

\subsection{Decentralized recommenders}

 In a context where the number of items and users grows by the minute, one of the main challenges of centralized recommenders remains their scalability. %
 To tackle this issue, decentralized approaches have been proposed for recommendation purposes in the context of matrix factorization~\cite{hegedus:2020:matrixgossip},
 or \ac{KNN}-based collaborative filtering~\cite{DBLP:conf/ipps/BoutetFGJK13} but also more generally for numerous machine learning problems~\cite{koloskova:2019:decentralizedsgd,ormandi:2013:gossiplearning,vanhaesebrouck:2017:decentralizedcl,nadiradze:2020:swarmsgd}.
 Most decentralized approaches rely on a gossip protocol to quickly disseminate information, typically model parameters in decentralized learning systems or user profiles in \ac{KNN}-based systems.

 In such a system, we assume that nodes are connected according to a specific topology such as a random graph. %
 Periodically, each node after having performed some local learning task, picks a number of neighbors in the topology %
 to forward them some information~\cite{DBLP:journals/tocs/JelasityVGKS07}. %
 This can be, for instance, the output of local learning tasks. %
 Relying on such a gossiping protocol enables the data or model to be disseminated in the network until convergence is reached.
 \rex{} relies on such a gossip protocol, which will be detailed in Section \ref{section:rex}.

\subsection{SGX}
\label{sec:back:sgx}

Since late 2015, Intel processors come with a hardware shielding subsystem called \acf{SGX}.
It consists of a user-level protection against any other process in the system, including higher-privileged ones that belong, for instance, to the \ac{OS} or the hypervisor.
This is achieved by automatic memory encryption, attestation, integrity and freshness guarantees ensured by hardware.

Applications that leverage this technology must be split into trusted and untrusted partitions.
While the former is limited in terms of instructions they can perform, like \ac{IO}, the latter is free to use the entire instruction set.
The reason for this constraint is that such instructions require the intervention of higher privileged (and untrusted) entities.
As a consequence, transitioning between trusted and untrusted modes entails context switches that involve cryptographic operations, memory copies and \ac{TLB} flushes~\cite{costan:2016:sgxexplained}, which incur high performance overheads.

From a software development perspective, transitions from the enclave to untrusted mode are made through \acp{ocall}, whereas the opposite is called \acfp{ecall}.
Conceptually, these are similar to \acp{RPC}, where functions and arguments are marshalled together and executed in a separate memory space.
Due to the limitation of executing \ac{IO} instructions from trusted code, we have to resort to proxying these operations through \acp{ocall}.
This makes it harder to port legacy applications and libraries in \ac{SGX} enclaves, as forbidden instructions have to be traced and replaced by such proxies.

\subsection{SGX remote attestation}
\label{sed:sgx:attestation}

Attestation is a crucial feature of \ac{SGX}.
It allows for other processes (or other enclaves) to be sure about what code is running inside a \emph{target enclave} (the one being attested) at initialization time.
Once trust is established, exchange of sensitive data can take place.

In a nutshell, the target enclave generates a \emph{report} that contains a hash (or \emph{measurement}) of its initial state (code, data and other attributes) computed by hardware upon the enclave initialization.
Such report can only be locally verified by another enclave running on the same processor, as it is signed with a key only known by the local platform.
In case the verifying node (verifier) is remote, a special platform enclave called \ac{QE} %
is in charge of verifying the target's \emph{report} and converting it into a \emph{quote}.
This, in turn, is signed with a private key before being sent to the verifier.
The verifier then checks this signature with the aid of another service, namely %
\ac{DCAP}, which finally confirms or refutes the authenticity of the signature.

\section{\rex{}}
\label{section:rex}

In this section, we provide a detailed description of \rex{}, our novel SGX-based decentralized algorithm. 
We first describe the establishment of trust between nodes (\ref{sec:rexattest}), the enclave execution of our protocol (\ref{sec:enclinter}) and \rex{}'s raw data sharing algorithm (\ref{sec:rawdata:share}). Finally, we discuss the parallelization aspects (\ref{sed:parallel}) and implementation details (\ref{rex:impl}) of \rex{}.

\subsection{\label{sec:rexattest}\rex{} attestation}

For designing \rex{}, we departed from classical decentralized learning algorithms.
Decentralized systems are typically composed of %
processes that share the same code, with no pre-established hierarchy among them.
\rex{} is no different in this regard.
We however enforce this feature with the help of the \ac{SGX} attestation protocol (Section \ref{sed:sgx:attestation}). %
In \rex, each pair of \ac{SGX} nodes must mutually attest themselves before exchanging sensitive data, regardless of when they join the system.
This gives the guarantee that all enclaves share the exact same initial code, practically nullifying the possibility of having rogue (or Byzantine) enclaves, as they cannot deviate from the expected behavior.

After a fruitful attestation, each node is convinced about the integrity of each other's initial code and data segments.
In addition, a shared secret must be established for confidential communication.
In order to obtain this key, we take advantage of the \emph{user data} field in the quote, which is filled with the public key of a \ac{ECDH} scheme~\cite{strangio:2005:ecdh}.
Once attestation is confirmed, the other node's public key that piggybacked the quote is combined with the local private key for obtaining the shared secret. 

At this point, we have a confirmation that the other node runs in a safe and genuine \ac{SGX} platform, apart from having established a symmetric key for encrypted communication.
We however do not yet know \emph{what code} that node is running.
This is achieved by comparing the \emph{measurement} within the quote to an expected value.
In \rex{}, we require all nodes to run the exact same code, so that this expected value must be equal to the checker's own measurement.
If we wanted to allow enclaves with different code-bases, the distinct measurements would have to be either hard-coded in the enclave binary or somehow provided from trusted sources~\cite{gregor:2020:palaemon}, increasing the complexity of the attestation procedure~\cite{chen:2022:mage}.

\subsection{\label{sec:enclinter}Enclave interface and \rex protocol}

\definecolor{sgx}{HTML}{A6D4B6}                                                 
\tcbset{                                                                        
	title=Enclaved,                                                             
	arc=2mm,                                                                    
	boxsep=0.6mm,                                                               
	top=0.6mm,                                                                  
	bottom=0.2mm,                                                               
	left=5mm,                                                                   
	text width=\linewidth-15mm,                                                  
	enlarge left by=-5.8mm,                                                     
	colframe=sgx,                                                               
	coltitle=black,                                                             
	boxrule=0.2mm,                                                              
	halign title=right,                                                         
}
\begin{figure*}
	\noindent
	\begin{minipage}{0.21\textwidth}
	{\begingroup
		\removelatexerror
		\begin{algorithm}[H]
			\caption{\footnotesize\label{alg:untrusted}Untrusted code, responsible for the bootstrap of \rex{} and \ac{IO} operations}
			\notsotiny
			\Procedure{initialize}{ \label{algo1:init}
				$\func{read\_dataset}()$\;
				$\func{start\_network}()$\;
				$\func{ecall\_init}(\mathit{arguments})$\;\label{algo1:ecall:init}
			}
			\Procedure{on\_receive}{ \label{algo1:receive}
				\Input{blob}
				$\func{ecall\_input}(blob)$\; \label{algo1:ecall:input}
			}
			\Procedure{ocall\_send}{ \label{algo1:ocall}
				\Input{destination\\ blob}
				$\func{send}(\mathit{destination}, \mathit{blob})$\; \label{algo1:send}
			}
		\end{algorithm}
		\endgroup}
	\end{minipage}
\hspace{0.01\textwidth}
	\begin{minipage}{0.77\textwidth}
	{\begingroup
	\removelatexerror
	\begin{algorithm}[H]
	\caption{\footnotesize\label{alg:trusted}Trusted code that runs inside \ac{SGX} enclaves. It concerns both the attestation and \rex{} protocols}
	\begin{multicols}{2}
	\notsotiny 
	\Procedure{ecall\_init}{ \label{algo2:ecall:init}
		\Input{args}
		\nonl \begin{tcolorbox}
			\nextnr $\textrm{local\_train\_data}, \textrm{local\_test\_data}\leftarrow\func{extract}(\mathit{args})$\;
			\nextnr $\func{initialize\_data\_structures}(\mathit{args}$)\;
			\nextnr $\func{rex\_protocol}(\varnothing,\varnothing)$ \tcp*{epoch 0} \label{algo2:call:rex}
		\end{tcolorbox}
	}
	\Procedure{ecall\_input}{ \label{algo2:ecall:input}
		\Input{blob}
		\nonl \begin{tcolorbox}
			\nextnr $\textrm{src, ciphertext}\leftarrow\func{extract}(\mathit{blob})$\;
			\nextnr \If{$\func{attested}(\mathit{src})$}{
				\nextnr $\textrm{shared\_key}\leftarrow\func{get\_shared\_key}(\mathit{src})$\;
				\nextnr $\textrm{data}\leftarrow\func{decrypt}(\mathit{shared\_key}, \mathit{ciphertext})$\;
				\nextnr $\func{rex\_protocol}(\mathit{src}, \mathit{data})$
			}\Else{
				\nextnr $\func{attestation\_protocol}(\mathit{src})$ \label{algo2:call:attest}
			}
		\end{tcolorbox}
	}

	\Procedure{rex\_protocol}{ \label{algo2:rex:protocol}
		\Input{src\\ data}
		\nonl \begin{tcolorbox}
			\nextnr \If{$\func{ready\_to\_train}(\mathit{src},\mathit{data})$}{ \label{algo2:check:ready}
				\nextnr $\textrm{alien\_model, alien\_train\_data}\leftarrow\func{extract}(\mathit{data})$\;\label{algo2:get:data}
				\nextnr $\textrm{local\_model}.\func{merge}(\mathit{alien\_model})$\; \label{algo2:merge1}
				\nextnr $\textrm{local\_train\_data}.\func{append}(\mathit{alien\_train\_data})$\; \label{algo2:merge2}
				\nextnr $\textrm{local\_model}.\func{train}(\mathit{local\_train\_data})$\; \label{algo2:train}
				\nextnr $\textrm{shareable\_data}\leftarrow\func{sample}(\mathit{local\_train\_data})$\; \label{algo2:share1}
				\nextnr $\textrm{shareable\_model}\leftarrow\func{get\_model}(\mathit{local\_model})$\;
				\nextnr $\func{share}(\mathit{shareable\_data},\mathit{shareable\_model})$\;\label{algo2:share2}
				\nextnr $\textrm{local\_model}.\func{test}(\mathit{local\_test\_data})$\label{algo2:test}
			}
		\end{tcolorbox}
	}
	\end{multicols}
	\end{algorithm}	
	\endgroup}
	\end{minipage}

\end{figure*}
 
In \rex{}, we restrict the \ac{TCB}, \ie, the amount of code that runs within enclaves, to the strict minimum, so as to reduce the chances of having software bugs and vulnerabilities, which grow with the amount of lines of code.
The \ac{TCB} consists of the C++ \ac{STL} provided in the Intel \ac{SGX} \ac{SDK} and libraries that do not need \ac{IO} (json serialization and linear algebra), whereas disk and network operations are kept in untrusted mode.

Once attested, \rex{} nodes execute a typical event-based protocol that collects notifications from their neighbor nodes in the communication graph and perform specific tasks depending on a determined set of application-specific conditions.
These tasks, in turn, may generate more events to be shared with fellow nodes.
The high-level design of \rex{} is summarized in Algorithms \ref{alg:untrusted} and \ref{alg:trusted}.
\Cref{alg:untrusted} lists the procedures executed in untrusted mode, \ie, those related to the bootstrap and \ac{IO}, whereas \Cref{alg:trusted} presents the internal enclave structure.

At initialization, \rex{} reads the input dataset, starts the network and initializes the enclave (\Cref{alg:untrusted}, lines \ref{algo1:init}-\ref{algo1:ecall:init}).
Upon receiving messages from the network, the untrusted code relays them to the enclave (\Cref{alg:untrusted}, lines \ref{algo1:receive}-\ref{algo1:ecall:input}).
No privacy threat happens here as only attestation messages, which are not privacy-sensitive, are exchanged in clear text.
Any attempt of an attacker to forge attestation messages would fail as it does not have access to secrets protected in the trusted environment. 
In the opposite direction, \ie, for %
calls made from inside the enclave, the untrusted code relays encrypted output data to the network interface (\Cref{alg:untrusted}, lines \ref{algo1:ocall}-\ref{algo1:send}).

There are two entry points to the enclave code: at initialization (\emph{ecall\_init}) and when a message arrives (\emph{ecall\_input}).
The enclave bootstrap (\Cref{alg:trusted}, lines \ref{algo2:ecall:init}-\ref{algo2:call:rex}) consists of copying the local partition of the dataset into protected memory, initializing data structures and triggering the first training on the initial data (epoch 0).

Upon reception of a message (\Cref{alg:trusted}, lines \ref{algo2:ecall:input}-\ref{algo2:call:attest}), its source is identified.
Along with the sender identifier, there is possibly a ciphertext that needs to be decrypted.
In case the attestation procedure has already been successfully completed, a secret shared key, which is only accessible within the enclave, must have been established with the source node, in which case the message is deciphered and forwarded to the subroutine responsible for the \rex{} protocol.
Otherwise, the procedure that takes care of the attestation is called to manage the recognition of the sender.

When \emph{rex\_protocol} is called, it checks whether it can perform a training iteration (\Cref{alg:trusted}, line \ref{algo2:check:ready}).
This happens either in the first training on the local initial data (\ie, $\mathit{src}=\varnothing$ and $\mathit{data}=\varnothing$) or when it has received a message (possibly empty) from all its neighbors.
In case one of these conditions is met, raw data and model (possibly empty) are extracted from the input data, and a series of operations take place (\Cref{alg:trusted}, lines \ref{algo2:get:data}-\ref{algo2:test}). %
We classify them into 4 steps:
\begin{itemize}
	\item \textbf{merge} (lines \ref{algo2:merge1}-\ref{algo2:merge2}): 
	\begin{itemize}
		\item if $\mathit{alien\_model}$ is not empty, it is merged with the local model (see Section \ref{sec:rawdata:share}).
		\item if $\mathit{alien\_train\_data}$ is not empty, all non-duplicate data items are appended to the local training data store.
	\end{itemize}
	\item \textbf{train} (line \ref{algo2:train}): \ac{SGD} iterations are performed on the local training data in order to improve the local model. %
	\item \textbf{share} (lines \ref{algo2:share1}-\ref{algo2:share2}):
	\begin{itemize}
		\item if raw data sharing is activated, the local data store is sampled and the selected data items are shared with neighbors according to the sharing algorithm in place.
		\item if model sharing is on, the local model is shared with neighbors following the selected sharing algorithm.
	\end{itemize}
	\item \textbf{test} (line \ref{algo2:test}): predictions are made for data items contained in $\mathit{local\_test\_data}$, which was not used for training, and then compared to the ground truth in order to verify the quality of the current model.
\end{itemize}

\subsection{Raw data sharing}
\label{sec:rawdata:share}

\rex{} speeds up convergence by sharing raw data with neighbors as opposed to FL and DLS that share model parameters.
The amount of data is parametrizable (as a program argument) and randomly selected from the local raw data store (\Cref{alg:trusted}, line \ref{algo2:share1}), which is kept inside protected memory.

We support two algorithms which determine the set of neighbors that will receive the raw data: either a random one (\acs{RMW}) or all of them (D-PSGD).
This is inspired by the way models are shared in two decentralized learning schemes, that we describe next.

\subsubsection{\Ac{RMW}}
In \ac{RMW}, or Gossip learning~\cite{ormandi:2013:gossiplearning}, each node randomly selects one of its neighbors to send its current model.
Upon receiving a model, a node averages it with its own and improves the model by training upon its local data.
When \ac{RMW} is active, \rex{} sends the raw data (instead of the model) to the same randomly selected neighbor.

\subsubsection{Decentralized parallel SGD (D-PSGD)}\label{sec:dpsgd}
In this approach~\cite{lian:2017:dpsgd}, each node sends its model to all neighbors.
Along with the model, it also sends an integer corresponding to its degree (\ie, how many neighbors the sender has).
Upon receiving a model, the destination node merges it into its own through a weighted average based on the degrees (we use Metropolis-Hastings weights~\cite{xiao:2007:metropolis}). %
When a node has no embedding for a given user or item, we consider only those of its neighbors. %
When D-PSGD is active, \rex sends to all neighbors a sampling of the local raw data.

\subsection{\label{sed:parallel}Parallelization}
\rex{} parallelization can be seen from both multi- and single-node perspectives. Being decentralized, \rex{} nodes independently and concurrently run on multiple machines, periodically synchronizing with neighbors according to the network topology and sharing algorithms. Synchronization barriers are established when a given node receives a message from all its neighbors, thereby triggering subsequent iterations of the protocol. In the current version, we do not tackle fault tolerance aspects. We leave failure detection (\eg, heartbeats and timeouts) for future work.

Within a single node, \rex{} executes \emph{merge}-\emph{train}-\emph{share}-\emph{test} tasks sequentially. This is a requirement in model sharing schemes because each task depends on the result of the previous one. \rex could however execute \emph{share} in parallel with the other tasks, since raw data sharing is independent of computing steps. Although our implementation currently lacks this feature, it could only further increase the advantages of leveraging \rex{} due to increased parallelism.

\subsection{Implementation}\label{rex:impl}

We implemented \rex{} in about \num{4200} lines of code in C++. Additionally, we used Intel \ac{SGX} SSL~\cite{intel:2021:ssl} for cryptographic algorithms, a json library~\cite{lohmann:2021:json} for serialization during attestation, ZeroMQ~\cite{boccassi:2021:zmq} for communication, and Eigen~\cite{guennebaud:2021:eigen} for sparse matrices and vectors.

For the comparisons between \ac{SGX} and native (\ie, without SGX), we use the same code-base, but compiled with a different set of flags and linked to distinct libraries.
Specific calls to the \ac{SGX} \ac{SDK} or routines that only make sense in enclave mode (such as attestation) are either filtered out with pre-processor directives or replaced by alternatives.

Sharing data brings the question of how much to share in every epoch. 
We treat this as another hyperparameter and experiment with several different values in order to pick one that fits well according to accuracy versus time comparisons.
By selecting a random sample of required data points to share, we make the data sharing a stateless procedure.
Thus, nodes may send the same data points more than once, although the probability of duplicates decreases as the data size increases. 

Another point to note when nodes share data is the amount of processing time required in every epoch, which would continually increase with the growth of input training data.
This results in very long training times as the model begins to reach convergence.
We solve this by fixing the number of batches taken into account in every epoch to a predefined value.
Hence, each node takes a fixed number of \ac{SGD} steps in every epoch regardless of the data available.
As a result, the training time per epoch remains constant throughout the learning process. 
\section{Evaluation}
\label{section:eval}
We now present an extensive evaluation of \rex{}.
We start with simulated scenarios for \ac{DNN} (50 nodes) and \ac{MF} (610 and 50 nodes) to demonstrate the benefits of raw data sharing.
Afterwards, we focus on a distributed setup of 8 nodes running on 4 \ac{SGX} servers (2 processes per machine), where we evaluate the enclave overheads.
We describe next the experimental setup followed by the results and corresponding assessment.

\begin{figure*}
	\centering
	\includegraphics{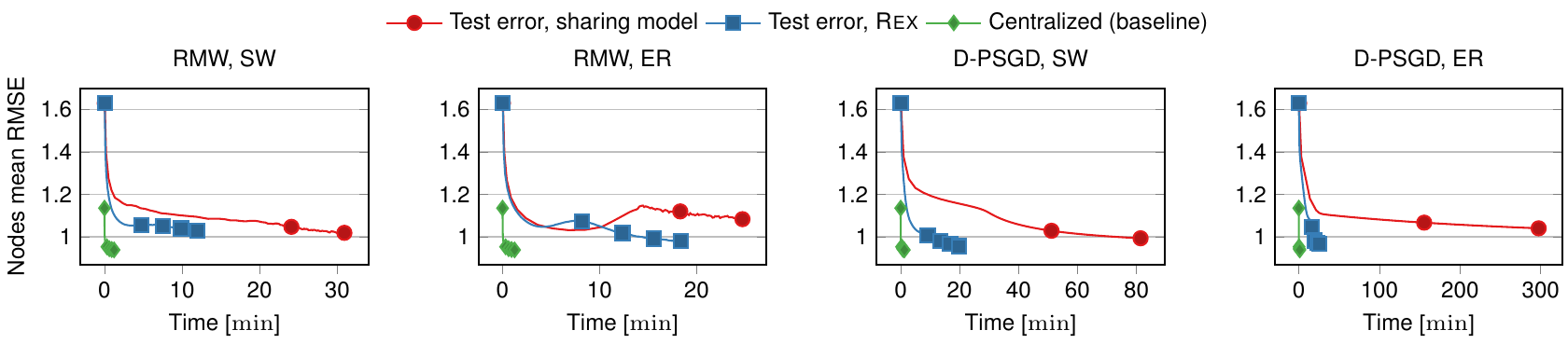}
	\vspace*{-4mm}
	\caption{\label{fig:1nodeperuser:time} One node per user | MF model. The figure charts evolution of test error with simulation elapsed time. \rex{} converges much faster than MS across all four cases, while the centralized baselines remains fastest as expected. Markers on the plots are spaced 50 epochs.}
\end{figure*}

\subsection{Experimental setup}
Apart from the distinct decentralized learning schemes we presented in Section \ref{sec:rawdata:share}, we use varied datasets in terms of size, and two network topologies. In this section, we also describe the metrics and experimental methodology we employed.

\subsubsection{Datasets}

We used MovieLens~\cite{grouplens:2021:movielens} datasets in our experiments, as shown in \Cref{tab:datasets}.
It consists of collections of movie ratings made by thousands of users in a website.
Users' ratings correspond to how much they appreciated a given movie, on a scale that ranges from 0.5 to 5.0, graphically represented by 5 stars which may be fully or partially filled.

{\footnotesize
	\begin{table}
		\centering
		\caption{\label{tab:datasets} Datasets.}
		\begin{tabular}{lrrrr}
			\toprule
			Dataset & {Ratings} & {Items} & {Users} & {Last updated} \\
			\midrule
			MovieLens Latest	& \num{100000} 	& \num{9000} 	& \num{610} 	& 2018\\
			MovieLens 25M* 		& \num{2249739}	& \num{28830}	& \num{15000}	& 2019\\
			\bottomrule
		\end{tabular} \\
	\vspace{1mm}
	*We capped the number of users (originally at \num{160000}), as our intent was to stay around the memory limits of our \ac{SGX} servers. Ratings and items correspond to this truncated dataset.
	\end{table}
}

\begin{figure*}
	\centering
	\includegraphics{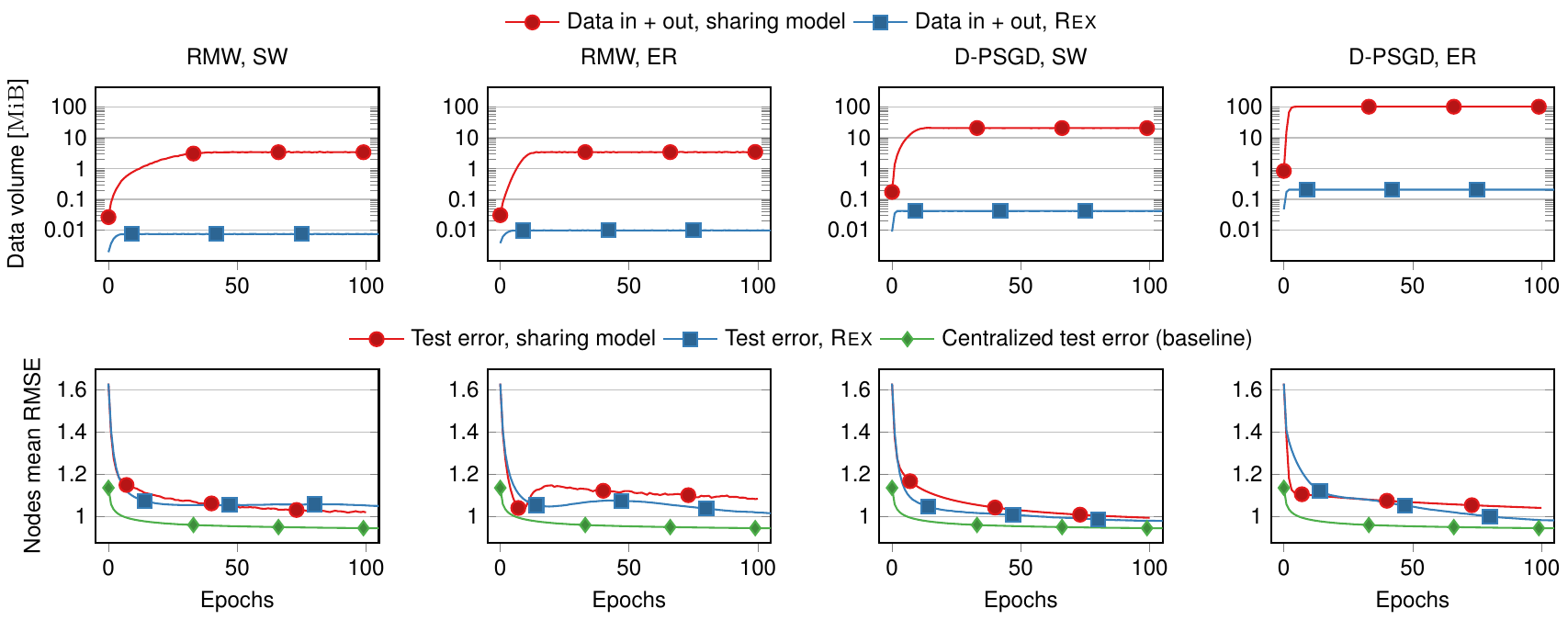}
	\vspace*{-4mm}
	\caption{\label{fig:1nodeperuser:epoch} One node per user | MF model. In correspondence to Figure \ref{fig:1nodeperuser:time}, this figure charts network usage and evolution of test error across epochs. Data exchanged in \rex{} is two orders of magnitude lower than MS across all four cases (Row-1). The test error in \rex{} evolves similarly to MS across epochs (Row-2) but each epoch runs significantly faster since only data is shared. Finally, the centralized baselines remains fastest in all cases.}
\end{figure*}

\subsubsection{Network topologies}

To assess our decentralized recommender system under different topologies, we chose Small World and random (Erd\H{o}s-R\'enyi), which we briefly describe next.

\paragraph{Small World}
This topology tries to mimic the relations that happen in real situations (\eg, social networks), where nodes are connected to small groups, out of which some may have far-fetched connections~\cite{SW}, according to the topological distance in the network. Each node has then close connections and a few far-fetched ones.
As a consequence, most nodes can reach each other in a small number of hops.
Technically, these graphs have low diameter and high clustering coefficient. In our experiments, we used a library called boost~\cite{siek:2021:boost:graph} to generate a SW topology taking as input  3 parameters: the size of the graph (610 and 50 here), the number of close connections (set to 6 in our experiments) and a probability of far-fetched connections (we set it to 3\%).

\paragraph{Erd\H{o}s-R\'enyi}
This topology consists of a random graph, where each edge is included in the graph with a given probability $p$~\cite{erdos:1959:randomgraphs}.
In comparison to Small World, these graphs may have larger diameters and lower clustering coefficients.
Although its construction mechanism can result in a disconnected graph (\ie, with multiple components), we ensure to make it connected by adding the missing edges. In our experiments, $p$ is set to 5\%.

\subsubsection{Machine Learning models}

\paragraph{\ac{MF}}
In the matrix factorization experiments, we split the dataset into train (70\%) and test (30\%) sets. %
We set the learning rate to $\eta=0.005$, the regularization parameter to $\lambda=0.1$ and the embedding dimension  to $k=10$. %
These values were obtained by several trials in the centralized setup.
The nodes share 300 data points per epoch.

\paragraph{\ac{DNN} model} %
For the DNN model, we have a setup with 50 nodes where each one holds the data of 12 or 13 users.
We use the Adam optimizer~\cite{diederik:2015:adam} with a learning rate of $\eta = 0.0001$, weight decay of 0.00001 and set the embedding dimension to $k = 20$. 
Following the embedding layer, the model has 4 hidden layers (linear + ReLU), dropout layers, and a final ReLU activation layer. 
The dropout rate for the embedding layer is 0.02 while for the first two hidden layers it is 0.15.
The described DNN model has \num{215001} model parameters in total.
Finally, in each epoch, the nodes share \num{40} data points. 

\subsubsection{Metrics}
We measure the benefits of \rex{} over 3 dimensions: training time, network traffic and test error as the \acf{RMSE}. %
Our goal is to show that raw data sharing renders better results than model sharing for all of them.
With respect to test error, it reaches a given value in a shorter amount of time.
We also  evaluate \rex{} on real \ac{SGX} servers, where we  measure the memory consumption as it represents a scarce resource in such environments.

\subsubsection{Methodology}
\label{sec:methodology}
We start by evaluating the scenario where each node holds the data of one user. %
This represents the situation where users initially have only their own data, \ie, what they produced.
Even though we use item ratings for recommendation, this situation would similarly apply, \eg, to text messages or pictures taken in a person's smartphone.

Next, we experiment with a setup %
where each node holds the data for several users, simulating a situation of distributed servers that are able to provide recommendations to these cohorts of users.
For example, \ac{SGX} servers in geographically-distributed data centers serving distinct clusters of users.

In our simulated experiments, we used servers with processor Intel Xeon E5-2630 v3 at 2.40GHz and and \SI{128}{\gibi\byte} RAM running Ubuntu 20.04.2 LTS kernel 5.4.0-72.
For the \ac{SGX} ones, we used 4 servers with processor Intel Xeon E-2288G CPU at \SI{3.70}{\giga\hertz} and \SI{64}{\gibi\byte} RAM running Ubuntu 18.04.4 LTS kernel 4.15.0-117 and the Intel \ac{SGX} \ac{SDK} v. 2.9.1.

\subsection{\rex{} versus model sharing} 
\label{sec:raw:versus:model}

We now present our experiments.
They are organized according to the experimental setting: one or multiple users per node, and those conducted on \ac{SGX} hardware.

\paragraph{One node, one user}

\Cref{fig:1nodeperuser:time} presents the evolution of test error with respect to the simulation elapsed time.
We use centralized execution as the baseline. %
Note that all scenarios converge to about the same error value, meaning that they are functionally equivalent.

Concerning the time to achieve a determined target error, we clearly observe that \rex{} is always better than sharing models.
To support this claim, we compile in \Cref{tab:timeratios} the values for an error target (chosen as the final value achieved by MS scheme), the times at which it was achieved and the ratio between timestamps.
\rex{} reaches speed-ups of up to $18.3 \times$  (D-PSGD, ER).
Additionally, we observe that D-PSGD is much slower than RMW. Whereas it took \SI{5}{\hour} to complete a simulation for D-PSGD ER, the longest RMW simulation took about \SI{30}{\minute} for the same number of epochs. This is due to the broadcasting nature of D-PSGD in contrast to the random neighbor unicast of RMW (Section \ref{sec:rawdata:share}).%

{\footnotesize
	\newcolumntype{g}{>{\columncolor{Gray}}r}
	\begin{table}
		\centering
		\caption{\label{tab:timeratios}  One node per user. Speedup in time achieved by \rex{} compared to model sharing (MS) for a given target error.}
		\vspace*{-3mm}
		\resizebox{\columnwidth}{!}{%
		\begin{tabular}{lSSSg}
			\toprule
			Setup		& {Error target}	& {\rex{} [\si{\minute}]}	& {MS [\si{\minute}]}	& \rex{} speed-up\\
			\midrule
			D-PSGD, ER	& 1.04		 	& 16.3	& 297.5	& $18.3\times$\\
			RMW, ER		& 1.08			& 2.1	& 24.7	& $11.5\times$\\
			D-PSGD, SW 	& 0.99			& 10.8	& 81.4	& $7.5\times$\\
			RMW, SW		& 1.03			& 12.0	& 27.4	& $2.3\times$\\
			\bottomrule
		\end{tabular}
		}
	\end{table}
}

The first line of charts in \Cref{fig:1nodeperuser:epoch} explains one reason why \rex{} achieves the same results in less time.
In terms of the volume of exchanged data, we observe that for all scenarios, sharing models was more than 2 orders of magnitude more expensive than \rex{}.
This happens because recommender systems are trained upon small data.
A raw data item is represented by a triplet containing the user and item identifications, along with the rating.
The model, on the other hand, is large.
In the case of \ac{MF}, a data item is associated to two feature vectors (or embeddings) related to the user and item in the triplet.
Each of these vectors alone is already larger than the data item to which they are associated.

\begin{figure}
	\centering
	\includegraphics{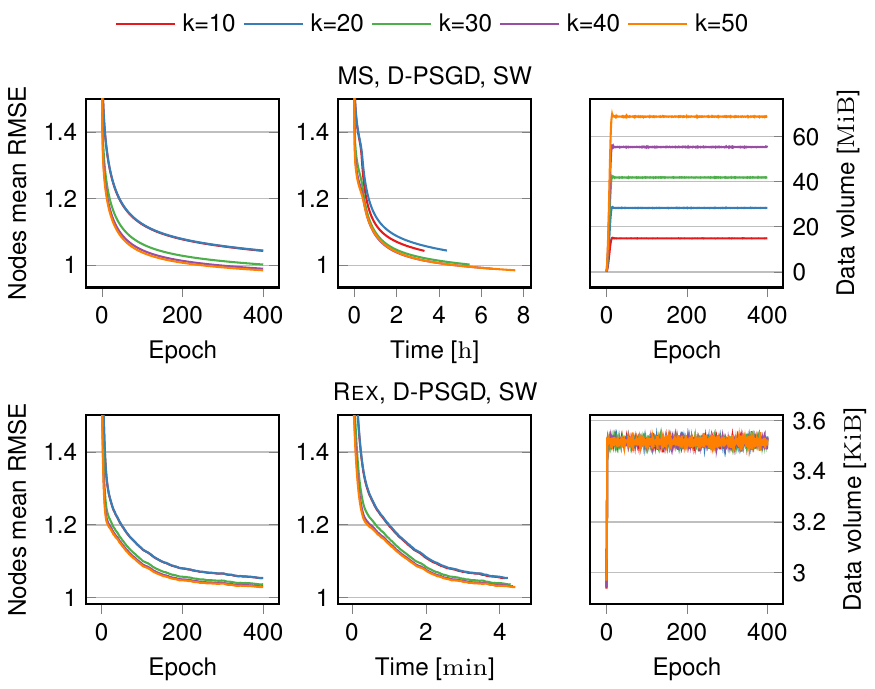}
	\vspace*{-4mm}
	\caption{\label{fig:varyK} Effect of varying feature vector size for D-PSGD, SW | MF model. All scenarios ran for fixed 400 epochs. Columns 1 and 2 chart test loss while column 3 charts data exchanged per node per round. For the MS case (row-1), increasing feature vector size provides little benefit in convergence time for the corresponding linear increase in network load. The impact on convergence of \rex{} (row-2) also remains little while the network load remains constant since only data is shared.}
\end{figure}

\begin{figure*}
	\centering
	\includegraphics{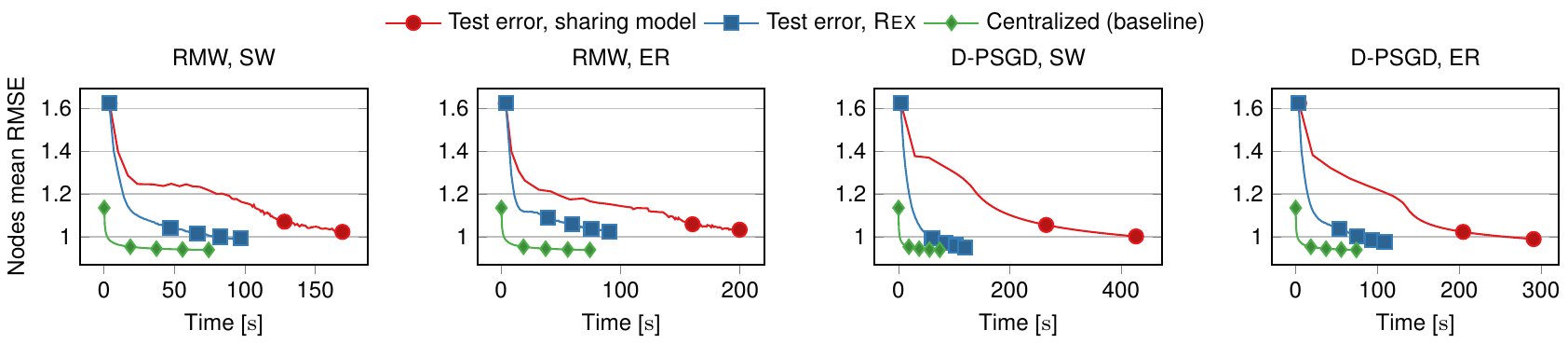}
	\vspace*{-4mm}
	\caption{\label{fig:my_label} Multiple users per node | MF model. The figure charts evolution of test error with simulation elapsed time. Similar to one node per user scenario, \rex{} converges much faster than MS across all four cases, while the centralized baselines remains fastest as expected. (Plots markers are spaced 50 epochs)}%
\end{figure*}

To evaluate the impact of the size of feature vectors, we ran the scenario with D-PSGD, SW for different lengths of embeddings and the equivalent \rex{} execution. Results are shown in \Cref{fig:varyK}. 
Each scenario is run for a fixed number of epochs (400).
As expected, \rex{} is not affected by feature vectors ($2^{nd}$ row, $3^{rd}$ column) in terms of network load because it does not share models. 
When models are shared ($1^{st}$ row), we observe that network load linearly increases at little benefit to convergence time upon increasing the size of feature vectors.
Thus, in our experiments, we set them to a fairly small size, equal to \num{10}. %
We found this to be a good compromise between having a reasonably accurate model and avoiding introducing bias towards our data sharing proposal by making models even bigger.

In terms of epochs, the charts in the second line of \Cref{fig:1nodeperuser:epoch} show that decentralized settings need more iterations in order to achieve the same error target as in the centralized equivalent.
This is inherent to their lack of global knowledge.
While the global model can uniformly improve for all the dataset at each iteration, decentralized alternatives can only count on local data plus the interactions with closest neighbors, thus delaying the progress of information dissemination.
In any case, even though \rex{} and model sharing take roughly the same number of epochs to converge, \rex{} is much better in terms of network and time. 

\paragraph{Multiple users per node}
\label{sec:eval:manyunsers}

Our following experiment tested our system in a second scenario: when the data of multiple users is initially partitioned across a number of servers (Section \ref{sec:methodology}).
In this setup, simulation times were much faster due to the fewer nodes through which information had to propagate.
We partitioned the ratings of the 610 users through 50 nodes and got similar results with respect to model and raw data sharing. This time, however, ratios were more modest. The results displayed on \Cref{fig:my_label} and 
\Cref{tab:timeratios:50nodes} summarizes them.
The reason why speedup is lower for multiple users per node is due to data concentration. As each node holds more data, we need less iterations to achieve a given target error, lowering the impact of network load, and hence \rex.

We then experimented with our \ac{DNN} recommender.
It was developed in \num{1495} lines of Python and uses PyTorch~\cite{paszke:2019:pytorch} for the \ac{DNN}, ZeroMQ~\cite{boccassi:2021:zmq} for communication and D-PSGD (Section \ref{sec:rawdata:share}) as sharing scheme.
\Cref{fig:dnn} displays the results. %

\ac{DNN}  results match the previous ones and show a lower epoch duration for \rex{} (\Cref{fig:dnn}\emph{(a)}), even though the difference is slightly smaller. %
Similarly, with respect to the amount of data exchanged,  we observe that volumes are orders of magnitude larger for model sharing (\Cref{fig:dnn}\emph{(b)}) in comparison to \rex.
Concerning the test error (\Cref{fig:dnn}\emph{(c)}), we observe that results vary according to the topology.
While small world (SW) achieves very similar results between the two sharing schemes, the random graph (ER) performs slightly worse for \rex{}, \ie, it achieves a larger error after a fixed number of epochs. %
We conjecture the reason to be related to the sparsity of the random graph, less connected than small world in this 50-node scenario. 
While MS encapsulates and propagates more information by training on entire local data, DS exchanges limited knowledge (contained only in the shared data points). %

{\begin{figure}
	\includegraphics{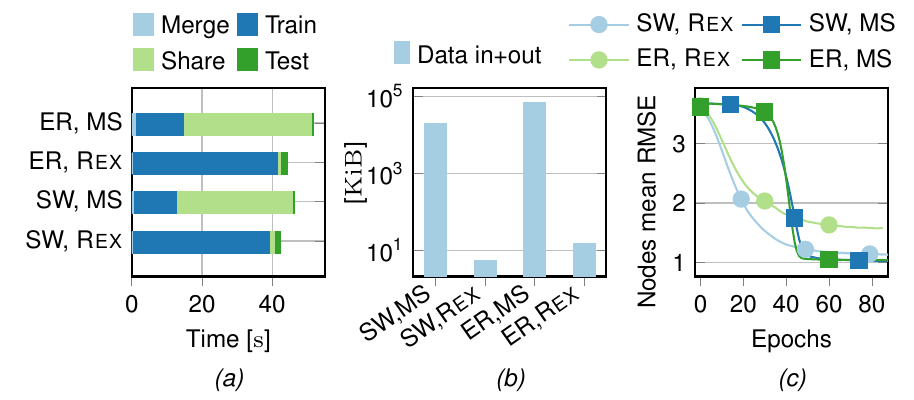}
	\vspace*{-8mm}
	\caption{\label{fig:dnn} Multiple users per node | DNN model. \emph{(a)} Time breakout of stages within an epoch (average across all nodes) - \rex{} is slightly faster. \emph{(b)} Data volume exchanged per epoch - \rex{} exchanges significantly less data than MS. \emph{(c)} Test error evolution per epochs - For SW, \rex{} converges faster than MS while achieving similar test error whereas for ER, \rex{} performs slightly worse.}
\end{figure}
}

{\footnotesize
	\setlength{\textfloatsep}{3pt}
	\newcolumntype{g}{>{\columncolor{Gray}}S}
	\begin{table}
		\centering
		\caption{\label{tab:timeratios:50nodes} Multiple users per node. Speed-up in time achieved by \rex compared to model sharing (MS) for a given target error.}
		\vspace*{-3mm}
		\begin{tabular}{lSSSg}
			\toprule
			Setup		& {Error target}	& {\rex{} [\si{\second}]} & {MS [\si{\second}]}	& {\rex{} speed-up}\\
			\midrule
			D-PSGD, ER	&0.99	&87.8	&292.5	&3.3$\times$\\
			RMW, ER		&1.03	&82.9	&200.6	&2.4$\times$\\
			D-PSGD, SW 	&1.00	&57.0	&430.4	&7.5$\times$\\
			RMW, SW		&1.02	&61.1	&170.1	&2.8$\times$\\
			\bottomrule
		\end{tabular}
	\end{table}
}

\begin{figure*}
	\centering
	\includegraphics{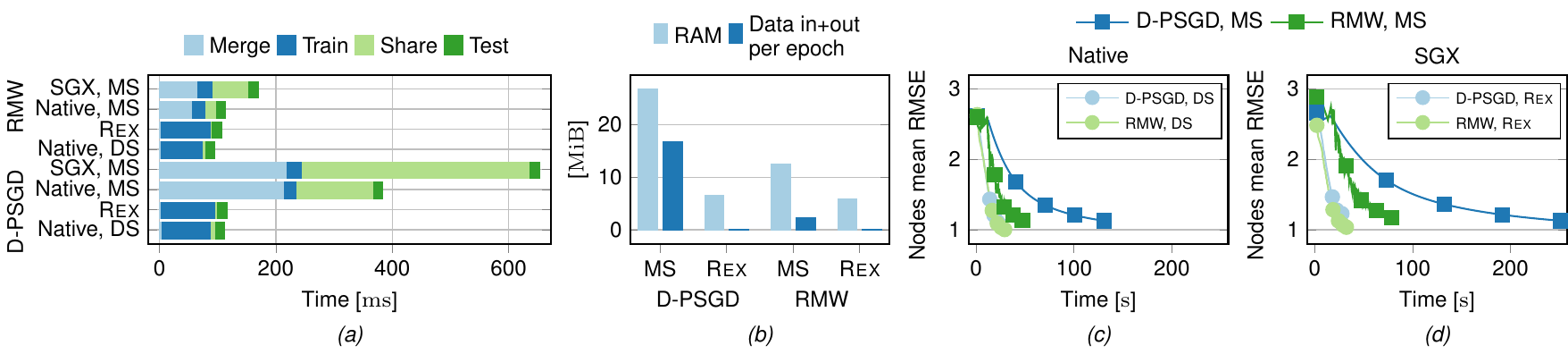}
	\vspace*{-4mm}
	\caption{\label{fig:time:break} Performance comparison with and without SGX for low memory usage (MovieLens Latest with 610 users) | MF model. \emph{(a)} Time breakdown of steps in an epoch - 
	Duration of merging and sharing is very low for \rex{} as compared to MS since only data is exchanged while the training duration remains high given the SGX procedures (\Cref{sec:back:sgx}). However, altogether \rex{} is faster than MS. The native equivalent runs faster for both DS and MS as expected. \emph{(b)} Memory and network usage - \rex{} exchanges much less data and requires less memory than MS. \emph{(c)} and \emph{(d)} Convergence speed (marker each 50 epochs) - \rex{} converges faster than MS similar to previous simulated experiments with very little overhead.}
\end{figure*}

\begin{figure*}
	\centering
	\includegraphics{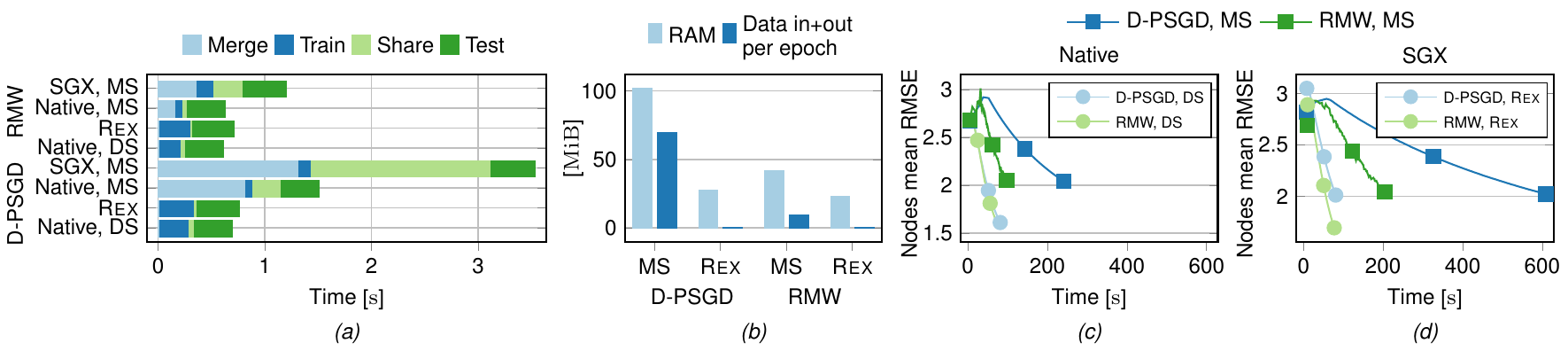}
	\vspace*{-4mm}
	\caption{\label{fig:beoynd:epc} Performance comparison with and without SGX for memory usage beyond EPC limit (MovieLens 25M with 15k users) | MF model. The observed trends for \emph{(a)} Time breakdown of steps in an epoch, \emph{(b)} Memory and network usage, \emph{(c)} and \emph{(d)} Convergence speed remain very similar to the scenario of low memory usage in Figure \ref{fig:time:break}.}
\end{figure*}

\subsection{SGX experiments}

Next, we measure \rex{} in SGX-capable machines in a distributed setup.
We used a 4-node network and ran 2 processes per machine in a fully connected setup, \ie, 8 nodes and 28 pair-wise connections.
Results are shown on Figs. \ref{fig:time:break} (low memory usage) and \ref{fig:beoynd:epc} (memory beyond EPC). %

In Figures \ref{fig:time:break}\emph{(a)} and \ref{fig:beoynd:epc}\emph{(a)}, we see the time breakdown according to each step of the distributed training process: merge, train, share and test.
The values correspond to the mean time that each step took per epoch across all nodes. %
We observe that sharing data (\rex and Native, DS) is always faster in comparison to exchanging models (MS).
The reason is the extra time needed for merging and sharing models.
As discussed in Section \ref{sec:rawdata:share}, the contributions of a group of neighbors are averaged together when models are exchanged.
While this procedure obviously takes some time, it is completely bypassed by \rex{}. %
Although we still need to check for duplicates, new data items are simply dumped into the local store with no further processing.
This is much faster than locating the relevant embeddings, attributing weights for neighbor contributions and performing the average.

When it comes to the duration of sharing, the difference is explained by the size of raw data and models (see Section \ref{sec:raw:versus:model}).
In Figures \ref{fig:time:break}\emph{(b)} and \ref{fig:beoynd:epc}\emph{(b)}, we see the average data volume exchanged per node per epoch. As previously shown, the data volume when sharing models is orders of magnitude higher, which justifies the extra overhead in sharing times for MS when compared to \rex{}.
Charts \emph{(c)} and \emph{(d)} of both Figures \ref{fig:time:break} and \ref{fig:beoynd:epc}, that present wall-clock time versus test error, were obtained in a non-simulated environment, \ie, with real network exchanges. They confirm the same pattern between MS and \rex{} found previously in our simulations.

\subsection{SGX and memory usage}

When putting into perspective the native (\ie, without SGX) and SGX experiments, we notice some slowdown in execution times for the latter.
Note that in native executions, data transmissions are in plaintext and there is no hardware protection.
Both raw data and models are therefore vulnerable in this case.
In the experiments of \Cref{fig:time:break}, we use the same dataset as in Section \ref{sec:raw:versus:model}, \ie, MovieLens with 610 users.
For those, the SGX overhead in terms of execution time varies from \SI{5}{\%} (\rex{}) %
to \SI{70.5}{\%} (model sharing).%

The reasons for the difference between \ac{SGX} and native executions lie in the intrinsic way enclaves are designed (see Section \ref{sec:back:sgx}), specially with respect to memory usage, transitions between the trusted and untrusted environments and all cryptographic and integrity operations involved in the process.
This is why the sharing step presents the biggest difference when we compare its times for \ac{SGX} and native, \ie, because it simultaneously involves \ac{IO}, cryptographic operations and intensive memory usage.

Interestingly enough, we consistently observed a puzzling exception to this pattern.
For \rex{}, the sharing step was slightly faster in the enclave execution.
We investigated the reason and found that the data sampling function was the source for this time difference.
The reason lies in the way memory is allocated.
While all enclave memory pages are obtained at initialization time, the native execution asks for more pages on-demand, therefore incurring in extra system calls to be serviced by the operating system during the sharing step, rather than in the bootstrap.
This behavior was one constraint of the first version of \ac{SGX} (\ie, the version of our machines), which was latter alleviated with the introduction of hardware support for dynamic memory allocation inside an enclave~\cite{xing:2016:memsgx2}.

In \Cref{fig:time:break}\emph{(b)}, along with network data volume exchanged, we see the amount of memory %
used throughout the execution.
The memory usage was measured with valgrind~\cite{nethercote:2007:valgrind} for the native implementation.
As we use identical codes in the native and SGX implementations (Section \ref{rex:impl}), the same measurements apply to the enclave.
The values correspond to the average heap usage sampled after the initial dataset input (when there is a peak of memory usage), after which it remains fairly constant. 
Since the dataset input happens in untrusted mode, this initial peak does not affect the \ac{SGX} execution and was therefore discarded for the sake of estimating the enclave memory usage.

In order to evaluate \rex{} in a more challenging setup, we took the MovieLens 25M dataset and limited the amount of users to \num{15000}.
This number was chosen because of the memory usage footprint it caused in our experiment.
More precisely, we wanted to have a condition where the \ac{EPC} is overcommitted.
The \ac{SGX} machines at our disposal have an \ac{EPC} of \SI{128}{\mebi\byte}, out of which only \SI{93.5}{\mebi\byte} are available for all enclaves running in each machine~\cite{vaucher:2018:sched}. 
In these experiments (\Cref{fig:beoynd:epc}), we reach more than twice the \ac{EPC} for D-PSGD MS and roughly the \ac{EPC} limit for RMW MS.
Both of them had considerable increase in the overhead when compared to the previous scenario. \Cref{tab:sgxoverhead} summarizes these results, which were obtained by comparing average time per epoch of \ac{SGX} over native.

{\footnotesize
\setlength{\textfloatsep}{3pt}
\begin{table}
	\centering
	\caption{\label{tab:sgxoverhead} The table presents overhead in execution time for \ac{SGX} w.r.t native. Presented alongside is the memory usage which explains the overhead. For the MS case, the overhead is significant (up to 135\%) but remains low for \rex{} (up to 17\%). }
	\resizebox{\columnwidth}{!}{%
	\begin{tabular}{lSSSS}
		\toprule
		& \multicolumn{2}{c}{610 users} & \multicolumn{2}{c}{\num{15000} users} \\
		Setup & {RAM [\si{\mebi\byte}]} & {Overh. [\si{\%}]} & {RAM [\si{\mebi\byte}]} & {Overh. [\si{\%}]}\\
		\midrule
		\rowcolor{Gray}
		RMW, \rex{} 	& 11.5 	& 14 	& 45.9 	& 17\\
		RMW, MS 	& 24.7	& 51	& 83.1	& 91\\
		\rowcolor{Gray}
		D-PSGD, \rex{} 	& 12.9 	& 5  	& 53.9 	& 8\\
		D-PSGD, MS 	& 53.6 	& 70 	& 204.0	& 135\\
		\bottomrule
	\end{tabular}
	}
\end{table}
}

\subsection{Discussion}

In this section, we recall \rex motivations and justify the absence of scalability experiments in this paper.
We address next the limitations of \rex, namely, \ac{SGX} vulnerabilities, poisoning attacks, vendor lock-in and memory constraints. Additionally, we list some future research avenues we would like to pursue.

\paragraph{Recap on \rex motivations}
With \rex, we demonstrated our three-dimensional goal of achieving at once: \emph{i}) privacy, with SGX enclaves; \emph{ii}) accuracy, shown through test loss in terms of \ac{RMSE}; and \emph{iii}) scalability, as a consequence of shorter convergence times and lower network usage (see \emph{b} below).
Our exciting results pave the way for further investigation on privacy-preserving raw data sharing in decentralized systems.
From a broader perspective, this would enable independent users to collaboratively train \ac{ML} models in a secure and scalable manner, as no dependence on centralized service providers is necessary.
This comes as an alternative to current recommender systems, which belong to giant tech companies who have access to private data of billions of users.

\paragraph{Scalability evaluation}
Although crucial in decentralized systems, scalability is a direct consequence of network topology and sharing algorithm~\cite{qiu:2004:bittorrent}. A fully connected topology scales poorly due to excessive connections, whereas RMW scales better than D-PSGD because of frugal network usage. This is however orthogonal to \rex, whose positive impact on scalability is secondary, \ie, it is a side-product of savings on network transfer. Consequently, in this paper, we chose to evaluate metrics directly impacted by the distinguishing design principles of \rex, \ie, data sharing and SGX. Given that we consistently achieve shorter convergence times and lower network usage, \rex can only improve decentralized recommender systems in terms of scalability. In addition, our resource limitations in terms of SGX hardware currently deters a proper scalability study of \rex.

\paragraph{SGX weaknesses} Hardware-enforced attestation at the application granularity is currently available only with SGX.
Although it guarantees that only trustworthy code runs inside enclaves, it does not prevent Byzantine users from subverting the system through poisoned input data, for instance.
Such attacks, along with those based on denial of service and side-channels, are not covered by the SGX threat model and therefore out the scope of this work.
Despite a few published attacks to SGX~\cite{vanbulck:2018:foreshadow,craciun:2020:sgxmalware,ragab:2021:crosstalk,chen:2021:voltpillager} and the mitigations that followed them, manufacturers keep investing and improving \acp{TEE}, which is a sign that such technology will keep evolving and hopefully will reach a maturity point when the feasibility of attacks will be very limited and swiftly neutralized.

\paragraph{Vendor lock-in and memory constraints}
Given that Intel Xeon platforms are \emph{full steam ahead} with \ac{SGX}~\cite{rao:2022:sgxrising}, we believe that \rex{} represents a viable solution for the future.
We hope however that multi-vendor groups, such as the \ac{TCG}~\cite{tcg:2022}, will eventually come up with standardized inter-operable \acp{TEE}, so that vendor lock-in will no longer be an issue.
With respect to memory limits,  Intel recently announced their new line of server processors with \ac{EPC} capacity of up to \SI{512}{\gibi\byte}, expandable to \SI{1}{\tebi\byte} when using two chips in one machine~\cite{intel:2021:3rdGenXeon}.
This will likely allow this technology to be widely used for memory-eager applications.  

Despite the technological infrastructure that enables \rex{}, the key takeaway of our proposal lies on the volume of raw data in perspective to models.
Apart from this, speed-ups can also come from the fact that sharing data can happen in parallel with the training, unlike model sharing which requires costlier aggregation and synchronization.

\paragraph{Recommenders versus other ML applications} Our choice on recommender systems was not incidental.
Among the reasons why we obtained considerable time and network gains is the high degree of sparsity in user profiles in such applications as well as the small size per data sample.
Because of that, we are interested in evaluating to which extent the same applies to other \ac{ML} applications (\eg, image classification, sentiment analysis, natural language processing).

Concerning model sharing, one could further reduce the amount of data that is exchanged as models by using gradient compression~\cite{koloskova:2019:compression,lin:2018:deepcompression,tang:2018:compression}.
Since recommendation systems are based on ratings that can take very few values (only 10 in the case of MovieLens, \ie, from 0.5 to 5.0 in steps of 0.5), data sharing in this area is also highly compressible.
In other domains, where data may already be compressed at the origin (\eg, pictures in JPG format), the compression rates would shrink. %
For these reasons, we leave for future work the assessment of the impact of gradient compression with respect to the choice of model or raw data sharing in decentralized training.
Moreover, data non-iidness is well-known to have a significant impact on the convergence of models in \ac{DL}.
We also plan on studying the impact of raw data sharing in the context of pathological non-iid datasets.

\section{Related work}
\label{section:related}

To the best of our knowledge, we are the first to use \ac{SGX} in a decentralized secure recommender system and leverage raw data sharing as a way to speed up training.
Nevertheless, privacy in recommender and decentralized learning systems was previously tackled, which we cover next.

\paragraph{Differential privacy and homormorphic encryption} 
Until the rise of \acp{TEE}, most practical approaches involved differential privacy or \ac{HE}. 
In this sense, Bellet et al.~\cite{bellet:2018:privatep2pml} propose differentially-private algorithms for decentralized systems, where a privacy budget $\epsilon$ is set in order to determine how much noise is added to data in order to prevent the disclosure of privacy-sensitive information. 
Boutet et al.~\cite{boutet:2016:ppdcf} propose the design of a decentralized recommender ensuring differential privacy through randomized protocols and a profile obfuscation mechanism. %
Danner et al.~\cite{danner:2018:decentrmlpriv}, in turn, combine heavy compression and a tree-based homomorphic encryption scheme to make a group of nodes jointly compute gradient sums in the context of a mini-batch \ac{SGD}. 
Nikolaenko et al.~\cite{10.1145/2508859.2516751} propose an approach to perform privacy-preserving matrix factorization through garbled circuits. %
Common to this line of work, one needs to handle the trade-off between accuracy, efficiency, and privacy. This is precisely what \rex avoids by using \acp{TEE}.

\paragraph{TEE-based decentralized systems} 
Using \ac{SGX} enclaves in decentralized systems was tackled in the domain of web-search relay networks.
Given the shielding and attestation capabilities of \ac{SGX}, enclaves were used to conceal user queries in such a way that adversaries are not able to inspect or subvert the behavior of relays.
SGX-Tor~\cite{kim:2017:torsgx} shows that this is achieved with low overheads.
In addition, Cyclosa~\cite{pires:2018:cyclosa} provides obfuscation mechanisms to frustrate Web-search engines attempts of re-identifying users.
These however do not involve gossip protocols to jointly compute \ac{ML} models as \rex does.

\paragraph{TEE and ML}
When it comes to \ac{ML}, many works use \acp{TEE} for security  and privacy~\cite{law:2020:xgboost,zheng:2019:helen, zhang:2021:citadel}.
Slalom~\cite{slalom:2019:tramer} uses the \ac{TEE} to keep the secrecy of linear layers in \acp{DNN} before leveraging hardware accelerators on concealed data.
Vessels~\cite{kim:2020:vessels} focuses on memory efficiency within enclaves.
In the context of \ac{FL}, PPFL~\cite{mo:2021:ppfl} uses SGX on the server-side and ARM TrustZone on worker nodes, whereas ShuffleFL~\cite{zhang:2021:shufflefl} protects the transmission of gradients with hardware enclaves along with a randomized scheme to prevent side-channel attacks. 
Unlike \rex, they are not target to decentralized systems. %
\section{Conclusion}
\label{sec:conclusion}

In this paper, we addressed privacy in distributed collaborative filtering systems and proposed the design, evaluation and implementation of \rex{}, the first SGX-based decentralized recommender. In the process, we have debunked the myth that there is an inescapable trade-off between accuracy or efficiency and privacy in collaborative filtering-based recommenders.  

Effectively, by leveraging \acp{TEE}, \rex{} enables raw data sharing among participants of decentralized systems without compromising user's privacy.
This contrasts with the traditional parameter sharing of federated learning and decentralized gossip-based approaches, which may yield privacy breaches.

We evaluated \rex{} across two network topologies, two model merging schemes and datasets of distinct sizes.
Moreover, we tried scenarios with one and multiple users per node.
We presented results for both native setting (without SGX) and in a (4 machine) distributed SGX environment.
Our results over all these settings consistently demonstrate that \rex{} improves up to $18.3 \times$ the training time, mostly due to network traffic which is significantly reduced.
Our implementation also demonstrates that the overhead of using TEEs remains negligible.  
At a time where SGX is starting to be available in major cloud providers, we believe that \rex{} is a credible approach to provide efficient, accurate recommendations in a wide range of applications without sacrificing on users' privacy. 
{	
	\printbibliography
}

\vspace{12pt}

\end{document}